\documentclass[preprint,floatfix,showpacs,groupedaddress,nofootinbib,superscriptaddress]{revtex4} 
\usepackage{amsmath,amssymb}
\usepackage{graphicx,color}

\newcommand{\norm}[1]{\left\|#1\right\|}
\newcommand{\be}{\begin{equation}}
\newcommand{\ee}{\end{equation}}

\newcommand{\p}{\partial}

\newcommand{\re}{{\rm e}}
\newcommand{\rd}{{\rm d}}

\newcommand{\EQ}{\begin{equation}}
\newcommand{\EN}{\end{equation}}

\newcommand{\n}{\mathbf{n}}
\newcommand{\Hi}{\mathcal{H}}
\DeclareMathOperator{\SU}{SU}
\newcommand{\g}{\mathfrak{g}}
\newcommand{\Hlow}{\Hi_{\mathrm{low}}}

\begin{document}
\title{On the relationship between  sigma models and spin chains } 
\author{D. Controzzi} 
\affiliation{ 
International School for Advanced Studies,
Via Beirut 4, 34014 Trieste, Italy} 
\author{E. Hawkins}
\affiliation{Department of Mathematics,
The University of Western Ontario,
London, Ontario N6A 5B7}

\begin{abstract}
We consider the two-dimensional $\rm O(3)$ non-linear sigma model with 
topological term using a lattice regularization introduced by 
Shankar and Read [Nucl.Phys. B336 (1990), 457], 
that is suitable for studying  the strong coupling regime. 
When this lattice model is quantized, the coefficient $\theta$ of the
topological term is quantized as 
$\theta=2\pi s$, with $s$ integer
or half-integer. 
We study in detail the relationship between 
the low energy behaviour
of this theory and the one-dimensional spin-$s$ Heisenberg model.
%This provides new evidence for the 
%validity of the Haldane mapping for general $s$.
We generalize the analysis to sigma models with other symmetries.
\end{abstract}

\maketitle

\section{Introduction}

The $\rm O(3)$ non--linear sigma model (NLSM) is a two--dimensional field 
theory  
of a 3-component, unit-vector field $n_\alpha$ ($\alpha=1,2,3$; 
$n^2=1$), with the Euclidean action given by 
\be
{\cal S} _{\theta} \,=\, \frac{1}{2g^2} \int \rd^2 x \left 
( \p _\mu \n \right )^2 + i \,\theta \,T \; ,
\label{o3.1}
\ee
where $g$ and $\theta$ are dimensionless coupling constants and 
\be
T\,=\,\frac{1}{8 \pi} \int \rd^2 x \; \epsilon^{\mu \nu} \;
\n\cdot(\p _\mu \n\times\p _\nu \n)
\ee
is an integer-valued topological term which measures the instanton number of a configuration \cite{belavin-polyakov};
it is well-defined provided that $\n$ 
converges to the same limit at infinity in all directions in the plane. 
In this case the physical space, as well as the target space, is effectively a
sphere, $S^2$, and the classical configurations are maps from $S^2$
to $S^2$; these are classified up to homotopy by the second homotopy group
$\pi_2(S^2)\cong\mathbb{Z}$. The topological term 
measures the homotopy class of such 
a configuration and is computed 
as the total area of the plane wrapped around the sphere, divided by $4\pi$.
Since $\theta$ enters the partition function as
$\re^{i\theta T}$ with $T$ integer, the model only depends on $\theta$ 
modulo $2\pi$.
%For $\theta=0$ 
The theory is known to be asymptotically free, 
i.e. it
behaves as a free theory at high energies but becomes strongly interacting at
small energies \cite{beta}. For $\theta=0$ the spectrum is 
known to be massive \cite{zam-zam-wiegmann}.
The topological term seems
to have no effect on the perturbation theory \cite{fendley.theta=pi}, 
nevertheless for $\theta=\pi$ the strong coupling behaviour changes 
completely. In fact, at some value of $g$, the flow will reach a 
fixed point
\cite{haldane1,haldane2,coadjoint.sigma,affleck,ah,affleck.lectures,fendley.theta=pi,shankar.read,merons}.    
Thus the excitations of the model at $\theta=\pi$ are massless
and it corresponds to the Renormalization Group (RG) 
flow between the $c=2$ Conformal Field Theory 
and the $\SU(2)_1$ Wess--Zumino--Witten (WZW) model at level $1$, with 
central 
charge $c = 1$ \cite{shankar.read,zam-zam.massless,ah}.
The two values $\theta = 0, \pi$ are the only ones 
for which the  action (\ref{o3.1}) is known to be 
integrable \cite{integrability} and the two-particle S-matrix was
suggested in
\cite{zam-zam-wiegmann} and \cite{zam-zam.massless} for $\theta=0$ and
$\theta=\pi$ respectively. The spectrum for generic values of $\theta$
was discussed in \cite{cm}.

There is general agreement  that the NLSM (\ref{o3.1}) with 
\be
\theta=2 \pi s
\label{theta}
\ee
describes the low energy behaviour of the one dimensional spin-$s$ 
quantum  Heisenberg model with antiferromagnetic interaction
($\mathcal{J}>0$)
\be
H_{\mathrm H}=\mathcal{J}\sum_{j=1}^N {\bf S}(j) \cdot {\bf S}(j+1),
\label{heisenberg}
\ee
where ${\bf S}(j)$ are spin-$s$ generators of $\SU(2)$. 
The relationship was first suggested by Haldane \cite{haldane1} 
who showed, using a
coherent state path integral description of the Heisenberg model, that for
large spin, $s\gg 1$, in 
the continuum limit this model can be mapped onto the NLSM with
$\theta$ given by (\ref{theta}). Other alternative derivations can be 
obtained
within the same approximation
\cite{haldane2,affleck.lectures,tsvelik}.
 
The identification \eqref{theta}, together with the periodicity of the sigma 
model in $\theta$, leads to the famous prediction that the
Heisenberg model is massive for integer spin and massless for half-integer
spins.
This prediction was confirmed by a recent study using non-Abelian
bosonization \cite{cabra}, and is in agreement with the 
exact solution of (\ref{heisenberg}) for
$s=\frac12$ \cite{bete} 
and numerical results on $s=1$ and $\frac32$ chains \cite{numerics-heis}. 
From the solution of the $s=\frac12$ chain one
also finds that the IR behaviour is given by the
$\SU(2)_1$ WZW model with a marginally irrelevant current-current
interaction \cite{heis.WZW}. 
%These results provide further evidence for the validity of the Haldane 
%mapping, 
%{\bf nevertheless, for general
%spin, one cannot be satisfied with a large spin approach.}

Shankar and Read \cite{shankar.read} suggested an interesting 
alternative analysis. They constructed a lattice
regularization of the NLSM with a topological term 
and showed that in a naive $g\to\infty$ limit
the $\theta=\pi$ model can be mapped to the $s=\frac12$ Heisenberg model. 
Specifically, as in the $\theta=0$ case \cite{hks},
this lattice Hamiltonian can be written as the sum of a kinetic and a 
potential term and, 
for $g$ ``large'', the latter can be treated as a perturbation. 
Shankar and Read were only interested in using this mapping to show that the 
NLSM with $\theta=\pi$ is massless and did not consider in detail the
relationship between the lattice sigma model and the Heisenberg model 
\eqref{heisenberg}. This is what we try to do in this paper. 
We use this analysis to gain some more insight into the 
validity of the Haldane mapping for general spin.
As we will see, the perturbative
approach can be easily generalized to any value of $\theta=2\pi s$
and it turns out to be related to the geometric quantization of $S^2$.
We carefully construct perturbation
theory up to second order and identify its limit of validity. 
The first order
result is the spin-$s$
Heisenberg model, while second order introduces next-nearest 
neighbor interactions (and quadrapole couplings); these are strongly reduced 
relative to the leading term by a very small ($s$ dependent) pre-factor.
At least for the $s=1/2$
case it is possible to show that this correction is irrelevant. 
Then condition  for the validity of the perturbative approach,
\eqref{weak.cond}, 
provides a {\em sufficient} condition for the validity of the
Haldane mapping.
In other words this approach shows that the low energy behaviour of the
spin-$s$ Heisenberg model is described by the O(3) NLSM with topological term 
$\theta$ given by (\ref{theta}) and $g$ that has to satisfy the condition
(\ref{weak.cond}). 
We also extend this approach to a more general class of sigma models.
These models have been considered in Ref.~\cite{salam} and, in
the context of the Quantum Hall Effect 
and disordered systems, in 
Ref.s~\cite{coadjoint.sigma,fendley2,fendley.theta=pi}

The paper is organized as follows. In the next section we recall the
Shankar-Read 
regularization procedure for the construction of a lattice regularized version 
of the NLSM and discuss its quantization in Section \ref{sec:quantum}. 
In Section \ref{perturbation} we consider perturbation theory and study its
implications for the Haldane mapping in Section \ref{sec:mapping}.
The extension to more general sigma models is done in 
Sec.~\ref{general.sigma}  and the
main results are discussed and summarized in the last section.

\section{Lattice regularization}

Of course the action (\ref{o3.1}) does not define a quantum theory unless one 
specifies how to regularize it. In fact, 
perturbation theory presents short distance divergences. 
In the asymptotically free regime ($g\ll 1$) standard renormalization
techniques can be applied  \cite{renormalization}. An alternative that is more
suitable for studying the strong coupling regime, is to put the theory on a
lattice \cite{hks,shankar.read} and then quantize it. We replace 
the spatial coordinate $x$ by a lattice with spacing $a$ and $N$ sites, but leave the 
Euclidean time $\tau$ continuous. The lattice regularization of the first term of \eqref{o3.1} is straightforward  and is given by
\be
\label{lattice.S0}
{\cal S} _0 \,=\, \frac{1}{g^2} \int \rd \tau \sum_{j=1}^N 
\left\{ \frac{a}2 \left( \p _\tau \n(j) \right )^2- \frac1a 
\n(j)\cdot \n(j+1) \right \},
\ee
plus an irrelevant constant. Clearly other regularizations, including e.g.,
next-nearest neighbor interactions, are possible. Nevertheless they
would give an irrelevant contribution at large distances. This regularization
correctly reproduces the spectrum of the NLSM with $\theta=0$ in the
strong coupling limit \cite{hks}.

Now, following Shankar and Read \cite{shankar.read}, 
we must regularize the topological term,
i.e., we have to construct a lattice term that approximates 
the area of the plane wrapped around the sphere. The lattice divides the
plane into strips and we can compute the topological term from the sum of the 
areas covered by each of these. Because of the boundary condition, each 
lattice point traces out a closed curve on the sphere. The area 
associated to the strip 
between $j$ and $j+1$ is simply the difference $\mathcal A(j+1)- \mathcal 
A(j)$ of the areas on the sphere 
enclosed by the curves traced out by $j$ and $j+1$.
Using Stokes' theorem, the area $\mathcal A(j)$ can be written as 
\be
{\cal A}(j) = 
\int \rd \tau \,{\bf A}(\n(j))\cdot  \frac{\rd {\bf n}(j)}{\rd \tau}
\label{area}
\ee
where ${\bf A}(\n)$ is chosen such that 
\be
\nabla \times \, {\bf A}=\n
\label{A.cond}
\ee
 along 
the unit sphere. It is not actually possible to choose such an $\mathbf A$ 
over the entire sphere, but it is enough to define it locally and in the 
quantum theory $\mathbf A$  plays the role of a background electromagnetic 
potential. Because of this, \eqref{area} may be incorrect by $\pm 4\pi$; 
however, this will eventually be irrelevant if $\theta$ is a multiple of $\pi$. We have
to keep this in mind when discussing some of the results below that follow
from the quantization of the theory.

The contribution to the topological term coming from the strip between two 
nearest sites is the difference between the associated areas: 
${\cal A}(j+1)-{\cal A}(j)$. This seems to be a very natural definition of the 
topological term. However, if we sum up these terms for all the strips, then 
everything cancels except a boundary term, thus not giving interesting physics. What Shankar and Read did was to carry out the 
summation only over every other strip and multiply by $2$, to get
\be 
T \approx \frac{1}{2\pi}
\int \rd \tau \sum_{j}^N (-)^j{\bf A}({\bf n}(j)) 
\cdot \frac{\rd {\bf n}(j)}{\rd \tau}.
\label{T.term}
\ee
In effect, the integrand in $T$ was rewritten as a finite difference on the 
lattice. Integration was then approximated by summation, but in a way that is 
not inverse to the finite difference used in the first step. Nevertheless,
one can see that for approximately 
continuous configurations, \eqref{T.term} indeed approximates the topological 
term. However, it is by no means the unique choice. Because the lattice has a 
different topology from the plane, it is not possible to systematically define 
a lattice-regularized topological term. At best, the ``topological'' term in 
a lattice model is merely analogous to the topological term in the continuum 
model. One justification for this is that, as we shall see below, it extends to a much more general analogy.

The lattice regularized form of the NLSM with topological term 
can then be written in the form
%\begin{multline}
\be
{\cal S} _{\theta} =  \int  \rd \tau \sum_{j=1}
\left \{  
\frac{a}{2g^2} \left ( \p _\tau \n(j) \right )^2 
%\right.  
%\qquad\qquad 
+ \frac{i\theta}{2\pi} (-)^j{\bf A}(\n(j))
\cdot \frac{\rd \n(j)}{\rd \tau}  - 
%\left. 
\frac{1}{a g^2}\n(j)\cdot \n(j+1) \right \}
\ee
%\end{multline}
where $a$ is the lattice spacing and the coupling constants in the rhs are all
dimensionless. If we redefine $\n(j)$ as $(-)^j\n(j)$, then we get rid of the
alternating sign in the second term and change the sign of the last term\footnote{$\mathbf A(j)$ can be defined independently at each site,
this allows us to use the same form at every site in the final expression.}.  
Now we can perform a Wick rotation and construct the Hamiltonian. The result 
is the lattice regularized version of the NLSM suggested by Shankar and
Read \cite{shankar.read}
\be
 H \,=\, \frac1a \sum_{j=1}^N 
\left \{ 
\frac{g^2}{2} L'^2(j)  + \frac{1}{g^2} \n(j)\cdot \n(j+1) \right \}
\label{o3.reg}
\ee
where 
\be
\mathbf L' := \n \times \left(\mbox{\boldmath$\pi$} - \frac{\theta}{2\pi}
\mathbf A\right)
\ee
and $\mbox{\boldmath$\pi$}(j)$ is the (formal) conjugate momentum to $\n(j)$. 
The first term in the sum is the kinetic energy of a charged particle moving on a unit sphere with a magnetic
monopole at the center, the strength of the monopole being given by
$\theta/2\pi$ \cite{monopole}.
We will
take this Hamiltonian as our starting point and use it to study the 
relationship
between the (continuum) NLSM (\ref{o3.1}) and the spin-$s$ Heisenberg model 
(\ref{heisenberg}). In particular we would like to verify the validity
of the Haldane mapping for general values of $s$.
Clearly the Heisenberg model
and the regularized sigma model (\ref{o3.reg}) 
are in general not equivalent at all scales. The latter is a
regularization of the NLSM that corresponds to the original theory  
(\ref{o3.1})
only at large distances (small energies).
If we now carry out the comparison at the
level of the two lattice models,
we need to show that part of the  
{\em low energy sector} of the lattice theory
(\ref{o3.reg}) 
can be mapped to the {\em low energy sector} of the Heisenberg model
(\ref{heisenberg}). 

\section{Quantum Model}
\label{sec:quantum}
In the lattice version of the NLSM \eqref{o3.reg}, the degrees of 
freedom at a single lattice site are equivalent to a charged particle moving 
on a sphere around a magnetic monopole \cite{monopole} with charge proportional to 
$\theta$.
The quantization of this model is now straightforward; it can be carried out 
systematically by using the formalism of geometric 
quantization \cite{GQ}, or one can 
use the known results for magnetic monopoles \cite{wu-yang}.
In the quantum model, the value of $\theta$ is 
quantized and  can only 
be a multiple of $\pi$. 
This corresponds to the quantization of magnetic monopole charge.
We can therefore write it as in (\ref{theta})
as $\theta=2\pi s$, where 
$s\in\frac12 \mathbb{Z}$. 
This seems in contrast with other quantization procedures
(\cite{affleck.lectures} and ref.'s therein) which work for other 
values of $\theta$, but it is a consequence of the necessity of assuming this in the derivation of the regularized topological term.
 
The subtlety of quantizing the Hamiltonian \eqref{o3.reg} is entirely with the first term. Quantizing the operator $\mathbf L'$ gives
\be
\mathbf L' := i\n \times \left(\nabla + i\frac{\theta}{2\pi}\mathbf A\right) .
\label{L.operator}
\ee
Because $\mathbf A$ satisfying Eq.~\eqref{A.cond} only exists over part of the sphere, this formula can only be applied over part of the sphere at one time. In order to define $\mathbf L'$ over the entire sphere, we must interpret $\frac{\theta}{2\pi}\mathbf A$ as a gauge potential, and $\nabla + i\frac{\theta}{2\pi}\mathbf A$ as a covariant derivative. Patching together gauge-equivalent operators, we find that the Hilbert space $\Hi_\mathrm{site}$ for a single lattice site consists of sections of a \emph{topologically nontrivial} line bundle over $S^2$. The curvature of the covariant derivative shows that the Chern character of the line bundle is proportional to $\theta$, but the Chern character is always integral, thus $\theta$ is quantized. If $\theta$ was not $2\pi s$, then it would not be possible to patch together consistently to define $\mathbf L'$.

Although it is not immediately apparent from Eq.~\eqref{L.operator}, this lattice model preserves the $\SU(2)$ symmetry of the original model. The operator $\mathbf L'$ is not quite the angular momentum, $\mathbf J$, as it does not satisfy the correct commutation relations. Instead, $\mathbf L' = \mathbf J - s\n$.  Because $L'^2 = J^2 - s^2$, we can redefine the Hamiltonian by an irrelevant constant, and write it most conveniently in terms of $J^2$.

Because the physics is the 
same for $-\theta$ as for $\theta$, we can assume without loss of generality 
that $s\geq 0$.
As a representation of 
$\SU(2)$, 
\be
\Hi_{\mathrm{site}} = (s) \oplus (s+1) \oplus (s+2) 
\oplus \dots
\ee
is the infinite direct sum of the spin $s$ representation, the spin $s+1$ representation, and so on.
In the case of $s=0$, this is just the Hilbert space of functions on 
$S^2$ and it is the direct sum of all integer-spin representations.
If we consider a finite lattice with $N$ sites, then the Hilbert space $\Hi$ for the model is the tensor product of $N$ copies of 
$\Hi_{\mathrm{site}}$,
\be
\Hi=\bigotimes_{j=1}^N\Hi_{\mathrm{site}}(j).
\ee

It is clear at this point that changing $\theta$ by $2\pi$ changes the model. This is in contrast to the formal partition function that we began with. We should expect that, if this is a valid regularization, then all the models with integer (respectively, half-integer) $s$ will have the same long-wavelength behaviour.

It is convenient to rewrite the quantum Hamiltonian as
\be
H = \frac{g^2}{a} K + \frac{1}{ag^2} V
\label{Hamiltonian2}
.\ee
%Here $a$ is the lattice spacing, and $g$ is the coupling constant appearing 
%in the $\sigma$-model Lagrangian \eqref{o3.1}. 
The ``kinetic'' operator is
\be
K = \tfrac12 \sum_{j} J^2(j),
\ee
while the
``potential'' operator is 
\be
V = \sum_j \n(j)\cdot \n(j+1),
\ee
where $\n(j)$ is the position vector on the unit sphere for the lattice site 
$j$. 
Note that $\theta$ does not appear explicitly in the Hamiltonian. Instead 
it 
enters into  the definitions of $\Hi_{\mathrm{site}}$  and the angular 
momentum  operators $\mathbf J(j)$. 

Let us consider a finite, circular lattice with $N$ sites and circumference 
$L=Na$. 
In this case, the operator $K$ 
is unbounded and has discrete spectrum, but the operator $V$ is bounded with 
continuous spectrum. 
In fact the angular momentum operator $J^2$ has spectrum
$\{l(l+1) \mid l=s,s+1,s+2, \ldots\}$. 
On the other hand
the inner product $\n(j)\cdot \n(j+1)$ 
takes values between $-1$ and $1$, and 
so has operator-norm $1$. The operator $V$ is the 
sum of $N$ such terms, hence 
it has norm $N$ and its spectrum is the closed interval from $-N$ to $N$.
The second term of the Hamiltonian \eqref{Hamiltonian2} 
is thus bounded, with norm
\be
\norm{\frac1{ag^2}V} =
\frac{N}{ag^2}.
\ee

The lowest eigenvalue of $K$ is $Ns(s+1)/2$. This contributes an 
(irrelevant) 
constant to the vacuum energy of the model. The eigenspace 
$\Hi_0\subset \Hi$,
associated to this eigenvalue, 
is the tensor 
product of a copy of the spin-$s$ representation for each 
lattice site. This is precisely the Hilbert space of the spin-$s$ Heisenberg 
model. 
The second eigenspace is spanned by states with spin $s+1$ 
at one site, and $s$ at all other sites; 
it is separated from the ground state
by a gap $\frac{(s+1)g^2}{a}$.
If this gap is more than 
the band width of the potential term, i.e.\ if
\be
\label{strong.cond}
(s+1)g^4 > 2N ,
\ee
then the gap must remain in the full theory \eqref{Hamiltonian2}. 
We notice that, at least on the bare level, inequality \eqref{strong.cond} is more likely satisfied for large spin\footnote{One should be careful about the dependence on $s$ of the behaviour of $g$ under Renormalization Group and of the possible renormalization of $s$ itself.}. In particular for a finite lattice and fixed $g$, this condition is always satisfied as $s\to \infty$.

\section{Perturbation theory}
\label{perturbation}
We shall analyze the Hamiltonian \eqref{Hamiltonian2} using perturbation theory about the strong coupling, $g\to\infty$ limit. Except for the overall factor of $g^2/a$, this means taking $K$ as the bare Hamiltonian, and treating $g^{-4}V$ as a perturbation. The expansion parameter is thus $g^{-4}$. Condition \eqref{strong.cond} is the sufficient condition for convergence of the perturbative series as well as for the validity of the perturbative approach. In fact, perturbation theory breaks down when distinct unperturbed 
energy levels intersect due to the perturbation.
We expect that for $g^4$ of the order $N/(s+1)$, 
the highest energy level of the 
lower band will intersect with a higher level, and a perturbative approach will 
no longer be rigorous. In the next section we will discuss the validity 
and meaning of a weaker condition.

If the inequality \eqref{strong.cond} is satisfied, 
the relevant part of the energy spectrum will spring entirely from the highly degenerate lowest eigenvalue of $K$.
As $g^{-1}$ increases from $0$, this lowest eigenvalue of $K$ splits into a discrete subspectrum of $H$, and we can follow these eigenvalues continuously, as long as \eqref{strong.cond} is satisfied. The eigenspaces of these eigenvalues give a well-defined subspace $\Hlow\subset\Hi$, with $\Hlow=\Hi_0$ for $g=\infty$. Using contour integrals, we can construct  orthogonal projections $\Pi$ and $\Pi_0$ on $\Hlow$ and $\Hi_0$ as,
\[
\Pi = \frac1{2\pi i} \oint \frac{\rd z}{z - K - g^{-4}V} 
\]
where the contour is (e.g.) the circle of radius $(s+1)N$ centered at $\frac12s(s+1)N$.
The remaining eigenstates have energy at least 
$\left (\frac{(s+1)g^2}{a} - \frac{N}{ag^2} \right)$
above the ground state and for energies smaller then this gap, 
such states become physically irrelevant. As we are interested in the low energy behaviour of this system, we should concentrate on $\Hlow$.

Because the Hamiltonians (for all $g$) have the same symmetries by $\SU(2)$ and lattice translations, it is possible to equivariantly identify $\Hlow$ and $\Hi_0$. If we choose some equivariant, unitary map $u : \Hi_0\to \Hlow$, then we can describe this low energy sector with an effective Hamiltonian,
\be
\label{H.eff}
H_{\mathrm{eff}} := u^* H u = u^* \left(\frac{g^2}{a} K + \frac{1}{ag^2} V\right) u
\ee
on $\Hi_0$.

If $u$ depends analytically on $g^{-4}$, then we can expand $H_{\mathrm{eff}}$ as a perturbative power series,
\be
\label{H.pert}
H_{\mathrm{eff}}=\frac{g^2}{a} h_0+\frac1{ag^2} h_1+\frac1{ag^6} h_2 +\ldots
\ee
It is important to keep in mind that this depends on some choice of $u$. The simplest choice of $u$ is $u = \Pi\, \Pi_0 (\Pi_0 \Pi\, \Pi_0)^{-1/2}$, although the resulting perturbation series may diverge prematurely. A more robust choice is defined by the conditions $u=\Pi_0$ for $g=\infty$ and $u^* \frac{\rd u}{\rd g} = 0$. However, the effective Hamiltonians for these two choices agree to second order, which is as far as we shall compute here.

Now consider the terms of the perturbative expansion \eqref{H.pert}. At order $0$, we have just the irrelevant constant $\frac{g^2}{a} h_0 = Ns(s+1)g^2/2a$.   The first order term is given by $h_1 = \Pi_0V\Pi_0$. The projection $\Pi_0$ onto $\Hi_0$ is the product of the projection onto the spin-$s$ representation $(s)\subset \Hi_{\mathrm{site}}$ for each site. For a function $f$ on the sphere, $\Pi_0 f \, \Pi_0$ is the Toeplitz operator of $f$, used in the geometric quantization of the sphere \cite{GQ}. In these terms, the operator $h_1$ is the sum of the products of Toeplitz operators of $\n$ at neighboring sites. The projection $\Pi_0$ is equivariant, so the Toeplitz operators $\Pi_0\n(j)\Pi_0$ must transform with spin $1$; consequently, they must be proportional to $\mathbf{S}(j)$, the angular momentum operator on $(s)$. The proportionality constant can be determined straightforwardly by computing one matrix element explicitly. This gives,
\[
\Pi_0 \n(j) \Pi_0 = \frac1{s+1} \mathbf S(j)
\]
From this, we see that the 
first order correction to the effective Hamiltonian is the Heisenberg Hamiltonian,
\be
\frac1{ag^2} h_1 = \frac1{ag^2} \Pi_0V\Pi_0 = \frac1{ag^2(s+1)^2} \sum_j \mathbf S(j)\cdot \mathbf S(j+1) .
\ee
%The norm of the further corrections is bounded by 
%\be
%\frac{N}{ag^6}
%.\ee

We can compute the second order correction explicitly. Let $P$ be a parametrix (approximate inverse) to $K - \frac12 Ns(s+1)$. Specifically, let $P$ act as the inverse on vectors orthogonal to $\Hi_0$, and as $0$ on $\Hi_0$. 
The second order correction to the effective Hamiltonian is given by 
\[
h_2 = -\Pi_0 V P V \Pi_0 .
\]
To write this explicitly, define the operator $S^{\alpha\beta}$ as the traceless, symmetric part of $S^\alpha S^\beta$; that is,
\be
\label{quadrapole}
S^{\alpha\beta} := \tfrac12\left(S^\alpha S^\beta + S^\beta S^\alpha\right) - \tfrac13 s(s+1) \delta^{\alpha\beta} .
\ee
In terms of this,
\begin{multline}
\label{h2}
h_2 = - \sum_j \left\{\tfrac{4s+1}{6(s+1)^3} 
- \tfrac{1}{2(s+1)^5} S^\alpha(j) S_\alpha(j+1)
+ \tfrac{8s+5}{2(2s+1)^2(s+1)^5} S^{\alpha\beta}(j) S_{\alpha\beta}(j+1) \right.\\
\left.+ \tfrac{2}{3(s+1)^4} S^\alpha(j)S_\alpha(j+2) 
+ \tfrac{2}{(2s+1)(s+1)^5} S^\alpha(j) S_{\alpha\beta}(j+1) S^\beta(j+2)\right\} .
\end{multline}
The computation is detailed in the appendix. We notice that the $1/a g^6$
correction is further dumped by a coefficient that contains higher powers of
$1/(s+1)$ and then will not modify significantly the long distance behaviour
of the theory even for finite values of $g$. 
Because $S^{\alpha\beta}=0$ for $s=\frac12$, \eqref{h2} 
simplifies considerably
\be
\label{half.spin}
h_2 = - \sum_j \left\{\tfrac{4}{27} 
- \tfrac{16}{243} S^\alpha(j) S_\alpha(j+1)
+ \tfrac{32}{243} S^\alpha(j)S_\alpha(j+2) \right\} .
\ee
We note that the sign of the next-nearest neighbor interaction is
negative and thus it is irrelevant at large distances for any value of $g$.
%that will be given by the Heisemberg model with the interaction renormalized
%by the second term in (\ref{half.spin}). 
This was not a priory obvious since a
next-nearest neighbor interaction with positive 
coupling constant ${\cal J}'>0.241
{\cal J}$ produce a frustrating effect that 
completely modify the large distance behaviour of the theory
\cite{eggert}. 

%It is possible that one can choose a different regularization of the 
%sigma model, including for instance
%next-nearest neighbor interactions, that will have the effect of cancelling out these 
%additional terms. We will not explore this possibility further. 

\section{The O(3) NLSM and the  Heisenberg model}
\label{sec:mapping}
Inequality \eqref{strong.cond} is a sufficient condition for the exact validity and 
convergence of the perturbative
approach.  The results of the previous section show that, if this is satisfied,
it is possible to rigorously map
the low energy sector of the regularized version of the O(3) NLSM onto a 
generalized Heisenberg model given by Eq.~\eqref{H.eff}.
The {\em full} spectrum of the Heisenberg model lies below the lowest 
energy gap of the lattice NLSM and is produced by the restriction of this
model to the Hilbert space ${\cal H}_{\mathrm{low}}$. 
Unfortunately, \eqref{strong.cond} is extremely restrictive and will always be 
violated in the thermodynamic limit as $N\to\infty$.
This condition comes from requiring that the
highest energy level of the lowest band does not intersect with a higher band.
However, we are not really interested in these highest energy levels, 
but in the lowest ones. 
In fact, we have to remember that our original aim was to check whether, 
at {\em low energies}, the
spin-$s$ Heisenberg model is described by the (continuum) 
NLSM (\ref{o3.1}) with $\theta=2\pi s$. This is much less than what we
have shown to be valid when (\ref{strong.cond}) is satisfied.
For our purposes
the mapping needs to be established only between the lowest energy
sectors of the two models. Thus if we focus only on the lowest energy
levels of the lattice NLSM,
it seems likely that these levels will not intersect levels coming
from an upper band and 
will continue 
to depend analytically on $g$, for \emph{much} lower values of $g$  than 
the one identified by the condition (\ref{strong.cond}). 
Thus conclusions about the low energy states obtained from
perturbation theory will 
continue to be valid even when (\ref{strong.cond}) is not satisfied.
In the rest of this section we will try to identify a condition for the
validity of the perturbative analysis for the lowest energy levels of the
lattice NLSM. Given the results of the analysis of the previous section,
this translates to a condition for the validity of the
mapping between the low energy sectors of the Heisenberg model and those of 
the NLSM.

Let us consider the lowest energy level of the  second band 
(i.e., the part of the spectrum originating from the second eigenvalue of
$K$). 
We expect the perturbative 
analysis to break down completely only when this level dives 
through the lower 
band and intersects with the ground state. The second band is not so 
different from 
the lowest band. It is dominated by states in which $N-1$ sites have spin $s$, and one site has spin $s+1$. The difference between this energy and the ground state should 
be given essentially by the effect of changing one site. 
There is clearly 
a positive contribution of $\frac{g^2}a (s+1)$ from increasing the spin. 
In addition, 
modifying one site will decrease the potential term by no
more than about 
$\frac{2}{ag^2}$ from the coupling of this site to its nearest neighbors. 
We thus estimate that this state will be separated from the ground state by an energy 
\be
\Delta_s(g)\gtrsim g^2 (s+1)/a-2/a g^2 .
\ee

If the theory is massless (the spin $s$ is half-integer), in the
thermodynamic limit, there exist many states with energies arbitrarily close to
the ground state. This suggests that as long as $\Delta_s(g)>0$ the
perturbative expansion for the lowest energy levels of the model remain valid
and provides the much weaker condition:
\be
\label{weak.cond}
g^4(s+1) \gtrsim 2 .
\ee
If the condition (\ref{weak.cond}) 
is satisfied, one expects that the behaviour of the 
low energy sector of the
lattice NLSM (which should be equivalent to that of its continuum limit) 
is correctly described by the perturbative analysis presented in the previous
section and 
is then equivalent to the low energy behaviour of the Heisenberg model.
%In addition, for 
%$s=1/2$ one can also see that the second order corrections 
%\eqref{half.spin} are irrelevant. 

If $s$ is integer the above reasoning has to be modified because the 
theory has a mass gap $M$, that translates to a finite correlation length
of the order of $1/M$, or $1/aM$ lattice sites. The most direct extension 
of the above results to take this into account is  to require
that $\Delta_s(g) > M$.  Now, we must keep in mind that the continuum 
approximation will break down if the correlation length is less than the 
lattice spacing $a$; this means that we must assume $aM<1$. Taking this into 
account when correcting \eqref{weak.cond}, we find that the correction is 
small. The weak condition \eqref{weak.cond} should only be trusted as an 
order of magnitude estimate, and as such it is unchanged in the massive case.
The same results can be obtained also in an alternative way. In fact the
presence of a finite correlation length implies that 
physics is insensitive to the total size of the 
lattice, as long as $N \gg 1/aM$. We can then use a relatively small 
lattice to test the relationship between the sigma and Heisenberg
models and replace the lattice length $N$ in
(\ref{strong.cond}) with the correlation length. 
This implies that conclusions based on perturbation theory will be 
valid if $(s+1)ag^4M > 1$. 

What do we learn from this analysis about the validity of the Haldane map for
general spin $s$? We started from the assumption that the lattice
model (\ref{o3.reg}) provides a valid lattice regularization for the O(3)
NLSM (\ref{o3.1}). There is no guarantee for this but it is a 
reasonable hypothesis previously used by other authors. 
This means that 
%at distances much larger then the lattice spacing, $a$, or
%equivalently 
at small energies, the continuum and lattice models have the same behaviour.
We showed that, if the requirement (\ref{weak.cond}) is satisfied, the low
energy behaviour of the lattice NLSM can be described within a perturbative
framework, carefully developed in the previous section up to second order.
The perturbative results show that when this approach is valid the low energy
sector of the lattice NLSM can be mapped onto the low energy sector of the
spin-$s$ Heisenberg model, where $s$ is related to the coefficient of the
topological term by (\ref{theta}).
Looking at things from the opposite direction we can say that the low energy
sector of the  spin-$s$
Heisenberg model is correctly described by the O(3) NLSM with $\theta$ given
by (\ref{theta}) and 
$g$ that satisfies the condition (\ref{weak.cond}). Since the 
sigma model is known to renormalize toward  strong coupling it is not necessary
that the inequality (\ref{weak.cond}) is satisfied on the bare level. Instead 
it can be satisfied at least at some intermediate scale. In principle, it is
then possible that the mapping will be valid independent of the value of the 
bare coupling
constant. Unfortunately the renormalization group behaviour of $g$ is not known
exactly, but, from what is known, 
it looks likely that condition (\ref{weak.cond})  will be satisfied at least 
if the bare value of $g$ is not too small.

%It is worth commenting on the spin dependence of the condition
%(\ref{weak.cond}). As it is easily seen the condition is more likely to be
%satisfied for large values of $s$, in agreement with what is found in the
%usual large $s$ approach. Since from other approaches 
%simplest case $s=1/2$. As we saw in the previous
%section, in this case the second order corrections are irrelevant and do not
%effect the low energy behaviour of the theory. Then In 
%addition we know that there is strong evidence for the validity of
%the mapping in the case $s=1/2$, as recalled in the
%introduction. Since, at least on the bare level, \eqref{weak.cond}
%is more easily satisfied 
%by higher spins, one can be fairly confident of the validity of the
%mapping for any half-integer spin. 

\section{More General Sigma Models}
\label{general.sigma}
Let us consider more general sigma models with a topological term. Space-time will still be $2$-dimensional, but instead of the sphere $S^2$, we allow a more general target space $\Sigma$. In order to preserve the symmetric character of the $\mathrm O(3)$ model, we assume that $\Sigma$ is a homogeneous space for some compact Lie group $G$ (i.e., it can be expressed as a coset space $G/H$). If we did not make this restriction, then the theory would have infinitely many coupling constants. We can also assume that $G$ is simple; if it were semisimple then we would essentially have a sum of two simpler theories. The case of $G$ Abelian is qualitatively different, so we don't consider that here either.

A classical configuration of the model is a map $\phi : \mathbb{R}^2\to\Sigma$ which converges to the same limit at $\infty$ in all directions. Because of this boundary condition, $\phi$ is effectively a map from the sphere $S^2$ to $\Sigma$. Such maps are classified topologically by the second homotopy group $\pi_2(\Sigma)$. 

In cases where $\pi_2(\Sigma)$ is a torsion group, such as $\mathbb{Z}_2$, it is possible to define a topological term for the partition function, but this cannot be written as a well-defined term in the action.

We are interested in adding to the action a topological term which can be written as a local integral. This is given by a $2$-form $\omega$ on $\Sigma$ which we integrate over the image of $S^2$. ``Topological'' means that this must be closed, $\rd\omega=0$. Without loss of generality, we can assume that $\omega$ is $G$-invariant. This (with the assumption that $G$ is simple) implies that $\Sigma$ is a \emph{coadjoint orbit} of $G$ (at least up to coverings).

The coadjoint space $\g^*$ is just the dual vector space to the Lie algebra $\g$ of $G$. (Because $G$ is simple, $\g^*$ and $\g$ are really the same for all intents and purposes.) There is a natural representation of $G$ on this space, and a coadjoint orbit is (by definition) the orbit of some point of $\g^*$ under the $G$ action. The form $\omega$ can be constructed canonically from the Lie algebra structure of $\g$.

The action for the more general sigma model is,
\be
\label{general.S}
\mathcal S_\theta = \frac1{2g^2} \int \rd^2x\, (\partial_\mu \phi)^2 + i\frac\theta{2\pi} \int \phi^*\omega
\ee
where $\phi(x) \in \Sigma \subset \g^*$ is a vector-valued function restricted to the given coadjoint orbit and $\phi^*\omega$ is the pull-back of $\omega$ to the space-time plane.

If $\Sigma$ is chosen to be an integral coadjoint orbit (see below) then the last integral in \eqref{general.S} is always a multiple of $2\pi$, and so changing $\theta$ by $2\pi$ does not change the partition function.
In the case of $S^2$, we take $\omega$ as $1/2$ the volume form so that $\int_{S^2}\omega = 2\pi$.

Aside from the simple case of $S^2$, this class of models includes the $\mathrm U(2N)/\mathrm U(N)\times \mathrm U(N)$ sigma models (with topological term) 
\cite{fendley2,fendley.theta=pi}
that have been proposed  
for modeling the transition between integer quantum Hall plateaus 
\cite{coadjoint.sigma}.

Because $\g^*$ is a vector space, the first part of the action can be regularized on a lattice exactly as in Eq.~\eqref{lattice.S0},
\be
{\cal S} _0 \,=\, \frac{1}{g^2} \int \rd \tau \sum_{j=1}^N 
\left\{ \frac{a}2 \left( \p _\tau \phi(j) \right )^2- \frac1a 
\phi(j)\cdot \phi(j+1) \right \}.
\ee

The derivation of the lattice version of the topological term also proceeds very much as above. The lattice divides the plane into strips. The topological integral can be written as a sum over the strips. The contribution from a strip is the difference $\mathcal A(j+1)-\mathcal A(j)$ of terms from the two lattice points bounding the strip. Again, if we include all the strips, then there is a massive cancellation and we are left with nothing interesting. So, we again approximate the integral by summing over half of the strips and multiplying by $2$.

This gives an alternating sum over the lattice sites.  Because of the boundary condition, $\phi(j)$ draws a closed curve on $\Sigma$; $\mathcal A(j)$ is defined (actually only defined modulo $2\pi$) by integrating $\omega$ over any disc which spans this curve. We can rewrite this as a line integral, 
\be
\mathcal A(j) = \int \frac{\rd \phi(j)}{\rd \tau} \cdot A \, \rd \tau
\ee
 using some $1$-form $A$ such that $\rd A = \omega$ 
(in a suitable neighborhood).
 
As in the derivation for $S^2$, we redefine the sign of every other $\phi(j)$, Wick rotate, construct the Hamiltonian, and quantize. The contribution of the topological term becomes a $\mathrm U(1)$ potential $\frac\theta\pi A$ on $\Sigma$. The Hilbert space for a site is thus the space of square-integrable sections of a line bundle with curvature $\frac\theta\pi\omega$. This line bundle and connection are $G$-equivariant.
The Hamiltonian is,
\be
H = \sum_j \left\{ \frac{g^2}{2a} J^2(j) + \frac1{ag^2} \phi(j) \cdot \phi(j+1)\right\}
\ee
where $J^2$ is the quadratic Casimir operator for the $G$-action.
%{\bf WHERE IS $\theta$?}

There is a correspondence between irreducible representations of $G$ and coadjoint orbits. This is given by the ``orbit method'' of constructing representations \cite{orbit.method}. Irreducible representations are classified by positive integral weights. The space of weights can be identified with a subspace of $\g^*$. Any coadjoint orbit is the orbit of a unique positive weight $\Lambda\in\g^*$; it is called integral if $\Lambda$ is an integral weight (and thus corresponds to a representation).

To construct the representation, we first construct a line bundle with curvature $\omega$ over the coadjoint orbit. The Hilbert space $(\Lambda)$ is the space of holomorphic sections of this line bundle. This can also be characterized as the subspace of sections with the lowest eigenvalue of the Laplacian --- or equivalently, the quadratic Casimir operator.

The Hilbert space $\Hi_{\mathrm{site}}$ for one site in this model is the space of square integrable sections of a line bundle with curvature $\frac\theta\pi\omega$. This is precisely the line bundle used in constructing the representation $(\frac\theta\pi\Lambda)$ by the orbit method. This only exists if $\frac\theta\pi\Lambda$ is integral. If $\Lambda$ is chosen carefully, then this means that $\theta$ is a multiple of $\pi$.

We can decompose $\Hi_{\mathrm{site}}$ into irreducible
representations, 
including $(\frac\theta\pi\Lambda)$ which is the lowest eigenspace of
$J^2$. 
The logic is the same as for the $\mathrm O(3)$ model. The lowest
eigenspace 
$\Hi_0$ of the kinetic operator $K$ is a tensor product with a factor
of 
$(\frac\theta\pi\Lambda)$ for each site. The gap between the two
lowest 
eigenvalues depends linearly on $\theta$.

If $g$ is sufficiently large, then we can identify a subspace $\Hlow\subset\Hi$ of low-energy eigenstates which is deformed from $\Hi_0$. Perturbation theory gives an effective Hamiltonian which is at first order the obvious generalization of the Heisenberg Hamiltonian \eqref{heisenberg}. The $\SU(2)$-representation $(s)$ is simply replaced by the $G$-representation $(\frac\theta\pi\Lambda)$, and the angular momentum operators are replaced with the $\g$-generators.

%The Haldane mapping can also be carried out here in the 
%large $\theta$ approximation. $\dots$

\section{Discussion}

In this paper we considered in some detail a lattice regularization
of the $\rm O(3)$ non-linear sigma model with topological term $i \theta T$ in
order to get some more insight on the validity of the Haldane map for general
values of the spin. 
The lattice NLSM  can be written as a sum of a kinetic and potential term
(cf.\ Eq.~(\ref{o3.reg}))
and the latter can be treated as a perturbation for $g$ sufficiently
large. When the model is quantized, $\theta$ is restricted to values 
$\theta=2 \pi
s$, with $s$ integer or half-integer. The Hilbert space associated to
the highly degenerate ground state of the unperturbed theory is a
tensor product of spin-$s$  representations of $\SU(2)$ at each lattice
site, i.e. exactly the Hilbert space of the Heisenberg model.
We carefully constructed perturbation theory up to second order and
identified its limits of validity. It turns out that, for the lowest energy
levels, the condition for the applicability of the perturbative approach is
given by (\ref{weak.cond}). If this condition is satisfied,
one can show that the low energy sector of the
lattice sigma model can be mapped onto the low energy sector of the
spin-$s$ Heisenberg model with effective coupling constant ${\cal
J}\approx 1/g^2(s+1)^2 a$.  Thus the condition of applicability of
perturbation theory also provides a sufficient condition for the validity of 
the Haldane mapping for general $s$.  
Due to  renormalization this inequality only needs to be satisfied  at
some intermediate scale. The condition (\ref{weak.cond}) 
depends on $s$ and shows explicitly that the 
mapping becomes more accurate (and valid on a wider range of energies) as $s$ 
increases.

We also extended the analysis to sigma models with a topological term
and a target space that is a homogeneous space of some compact Lie
group $G$. Again in this case, under some conditions,  it is possible to 
introduce a suitable
lattice regularization, and a mapping can be established between the low
energy sector of these more general lattice sigma models and the low energy
sector of generalized ``Heisenberg'' models where $\SU(2)$
representations are replaced by $G$-representations.

\section{Acknowledgments}
We would like to thank A. Nersesyan, A. Tsvelik, G. Mussardo for interesting
discussions. D.C would like to thank Brookhaven National Laboratory,
where this work was completed, for hospitality. For D.C. this work is
within the activity of the European Commission TMR program
HPRN-CT-2002-00325 (EUCLID) and Italian COFIN ``Teoria dei Campi,
Meccanica Statistica e Stati Elettronici''. 

\begin{appendix}
\section{Perturbation Details}
\label{appendix}
We now present the details of the computation of the expression \eqref{h2} for the second order correction $h_2$ to the effective Hamiltonian.

The second order term is $h_2 = -\Pi_0 V P V \Pi_0$. We can decompose the Hilbert space $\Hi$ into pieces based on the total spin at each site. $\Pi_0$ projects to $\Hi_0$, where every site has spin $s$. Multiplying a spin-$s$ state by the coordinate operators $n_\alpha(j)$ gives superpositions of spin $s$ and spin $s+1$ states at $j$. Applying $\n(j)\cdot\n(j+1)$ to a vector in $\Hi_0$ gives a superposition of a vector in $\Hi_0$, vectors with spin $s+1$ just at $j$ or $j+1$, and a vector with spin $s+1$ at $j$ and $j+1$. The parametrix $P$ vanishes on the first component, has eigenvalue $\frac1{s+1}$ on the second two, and eigenvalue $\frac1{2(s+1)}$ on the last.

\emph{A priori}, 
\[
h_2 = -\sum_{i,j=1}^N \Pi_0\, \n(i)\cdot\n(i+1)\, P\, \n(j)\cdot\n(j+1)\, \Pi_0 ,
\]
but the summands vanish unless $\{i,i+1\}$ overlaps with $\{j,j+1\}$. This naturally decomposes as 
\[
h_2 = \sum_j \left(h^{\mathrm{triple}}_j + h^{\mathrm{pair}}_j\right)
\]
where $h^{\mathrm{triple}}_j$ is the terms involving all three sites $j-1$, $j$, and $j+1$, and $h^{\mathrm{pair}}_j$ is the terms involving the two sites $j$ and $j+1$. 

In $h^{\mathrm{triple}}_j$, $P$ acts as $0$ for spin-$s$ at $j$, and as $\frac1{s+1}$ for spin $s+1$ at $j$. We can therefore replace $P$ with $\frac1{s+1}(1-\Pi_0)$ in this term,
\be
\label{triple}
\begin{aligned}
h^{\mathrm{triple}}_j
&= -\Pi_0[n^\alpha(j-1)\, n_\alpha(j)\, P n_\beta(j)\, n^\beta(j+1) + n_\beta(j)\, n^\beta(j+1)\, P n^\alpha(j-1)\, n_\alpha(j) ] \Pi_0 \\
&= \frac{-1}{(s+1)^3} S^\alpha\otimes \Pi_0\left(n_\alpha[1-\Pi_0]n_\beta + n_\beta[1-\Pi_0]n_\alpha\right)\Pi_0 \otimes S^\beta .
\end{aligned}
\ee
Some explicit computation gives the useful identity, $\Pi_0 n_\alpha n_\beta\Pi_0 = \frac2{(2s+1)(s+1)} S_{\alpha\beta} + \frac13\delta_{\alpha\beta}$. Using this, the  middle factor of \eqref{triple} becomes,
\[
\begin{aligned}
2\Pi_0n_\alpha n_\beta \Pi_0 - \tfrac1{(s+1)^2}(S_\alpha S_\beta + S_\beta S_\alpha) 
&= \tfrac4{(2s+1)(s+1)} S_{\alpha\beta} + \tfrac23 \delta_{\alpha\beta} - \tfrac2{(s+1)^2} S_{\alpha\beta} - \tfrac{2s}{3(s+1)} \delta_{\alpha\beta} \\
&= \tfrac{2}{(2s+1)(s+1)^2} S_{\alpha\beta} + \tfrac2{3(s+1)} \delta_{\alpha\beta} .
\end{aligned}
\]
This gives,
\[
h^{\mathrm{triple}}_j 
= \frac{-2}{(2s+1)(s+1)^5} S^\alpha \otimes S_{\alpha\beta} \otimes S^\beta - \frac2{3(s+1)^4} S^\alpha\otimes 1 \otimes S_\alpha ,
\]
which are the last two terms of Eq.~\eqref{h2}.

The other part,
\[
h^{\mathrm{pair}}_j = \Pi_0\, n^\alpha(j)n_\alpha(j+1)\, P\, n^\beta(j)n_\beta(j+1)\, \Pi_0
\]
is more complicated. There are only $4$ combinations of spins that can occur in the middle of this, therefore we can substitute
\[
(s+1)P \to \tfrac12(1\otimes 1)  + \tfrac12(\Pi_0\otimes 1 + 1\otimes\Pi_0) - \tfrac32 \Pi_0\otimes\Pi_0
\]
where $\Pi_0$ now denotes the projection onto $(s)$ at one site. $1\otimes 1$ gives
\[
\left(\tfrac{2}{(2s+1)(s+1)} S^{\alpha\beta} + \tfrac13\delta^{\alpha\beta}\right) \otimes \left(\tfrac{2}{(2s+1)(s+1)} S_{\alpha\beta} + \tfrac13\delta_{\alpha\beta}\right)
= \tfrac4{(2s+1)^2(s+1)^2} S^{\alpha\beta}\otimes S_{\alpha\beta} + \tfrac13 .
\]
$\Pi_0\otimes 1$ gives
\[
\begin{split}
\tfrac1{(s+1)^2} S^\alpha S^\beta \otimes \left(\tfrac{2}{(2s+1)(s+1)} S_{\alpha\beta} + \tfrac13\delta_{\alpha\beta}\right)
&= \left(\tfrac1{(s+1)^2} S^{\alpha\beta} + \tfrac{s}{3(s+1)} \delta^{\alpha\beta}\right)  \otimes \left(\tfrac{2}{(2s+1)(s+1)} S_{\alpha\beta} + \tfrac13\delta_{\alpha\beta}\right) \\
&= \tfrac{2}{(2s+1)(s+1)^3} S^{\alpha\beta}\otimes S_{\alpha\beta} + \tfrac{s}{3(s+1)} .
\end{split}
\]
$1\otimes\Pi_0$ gives the same thing. $\Pi_0\otimes\Pi_0$ gives
\[
\begin{split}
\tfrac1{(s+1)^4} S^\alpha S^\beta \otimes S_\alpha S_\beta 
&=  \tfrac1{(s+1)^4} \left(S^{\alpha\beta} + \tfrac{i}2\varepsilon^{\alpha\beta\gamma} S_\gamma + \tfrac{s(s+1)}3 \delta^{\alpha\beta}\right) \otimes \left(S_{\alpha\beta} + \tfrac{i}2\varepsilon_{\alpha\beta\delta} S^\delta + \tfrac{s(s+1)}3 \delta_{\alpha\beta}\right) \\
&= \tfrac1{(s+1)^4}S^{\alpha\beta}\otimes S_{\alpha\beta} - \tfrac1{2(s+1)^4} S^\alpha\otimes S_\alpha + \tfrac{s^2}{3(s+1)^2}
\end{split}
\]
This adds up to
\[
h^{\mathrm{pair}} = -\frac{8s+5}{2(2s+1)^2(s+1)^5} S^{\alpha\beta}\otimes S_{\alpha\beta} + \frac1{2(s+1)^5}S^\alpha\otimes S_\alpha - \frac{4s+1}{6(s+1)^3}
\]
which are the first 3 terms of \eqref{h2}.

\end{appendix}

\end{document}